\documentclass[preprint,aps,nofootinbib]{revtex4}
\usepackage[utf8]{inputenc}

\usepackage{amsmath, amsthm, amssymb}
\usepackage{latexsym}
\usepackage{epsfig}
\usepackage{epstopdf}
\usepackage{graphicx}
\usepackage{mathtools}
\usepackage{hyperref}
\usepackage{color}

\newcommand{\myref}{Ref.~}
\newcommand{\myrefs}{Refs.~}

\def\lsim{ \,\, \vcenter{\hbox{$\buildrel{\displaystyle <}\over\sim$}}
 \,\,}
\newcommand{\as}{\alpha_s}
\renewcommand{\v}[1]{ \ensuremath{  \vec {#1}_\perp }}

\newcommand{\QsP}{\ensuremath{{Q_{s}^{(P)}}}}
\newcommand{\QsA}{\ensuremath{{Q_{s}^{(A)}}}}

\begin{document}

\title{Saturation corrections to dilute-dense particle production and azimuthal correlations in the Color Glass Condensate}

\author{S.~Schlichting$^1$ and V.~Skokov$^{2,3}$}

\affiliation{$^1$Fakult\"at für Physik, Universit\"at Bielefeld, D-33615 Bielefeld, Germany\\
$^2$Department of Physics, North Carolina State University,
  Raleigh, NC 27695\\
  $^3$RIKEN-BNL Research Center, Brookhaven National Laboratory, Upton,
  NY 11973}

\date{\today}

\begin{abstract}
We perform a numerical study of higher order saturation corrections to the dilute-dense approximation for multi-particle production in high-energy hadronic collisions in the framework of the Color Glass Condensate. We compare semi-analytical results obtained by performing a leading order expansion in the dilute field of the projectile with numerical simulations of the full Classical Yang-Mills dynamics for a number of phenomenologically relevant observables. By varying the saturation momentum of the target and the projectile, we establish the regime of validity of the dilute-dense approximation and assess the magnitude and basic features of higher order saturation corrections.    
In particular,  we find that dilute-dense approximation faithfully reproduces dense-dense results if restricted to the range of its validity.

\end{abstract}

\maketitle
\section{Introduction}
\label{sec:Intro}
Describing multiple production of semi-hard particles in high-energy hadronic collisions is a challenging task, which in general is not well understood theoretically. At asymptotically high energies, the Color Glass Condensate effective theory (see e.g.~\cite{KovchegovLevin}) provides a viable approach to describe multi-particle production, including correlations between the produced particles. In its simplest form the colliding hadrons are approximated by two sheets of Classical Yang-Mills field $A\sim {\cal O}(g^{-1})$ with quantum corrections  suppressed by extra powers of strong coupling constant $g$. Particle production can be described by the classical gluon fields  after the collision  in the forward light cone. Within this framework, an analytical approach is possible when one of the objects can be considered as ``dilute''
$A\sim {\cal O}(g^{0})$. This allows one to perform the expansion in the measure of the diluteness, usually quantified by the projectile saturation momentum $\QsP$. Conversely, if both colliding objects are ``dense'' $A\sim {\cal O}(g^{-1})$, a full set of classical Yang-Mills equations has to be solved. The authors are not aware if even a distant possibility of having an analytical result exists in this case.  

\subsection{Particle production in classical approximation.}
In order to review the current theoretical  status of particle production in the saturation/CGC formalism,   
let us first consider single inclusive particle production. Schematically, the single inclusive particle gluon spectrum can be expressed as (see Ref.~\cite{Kovchegov:2018jun} for more details):  
 \begin{align}
   \frac{dN}{d^2 k dy } 
   = \frac{1}{\alpha_s} \, f \left( 
   \frac{\QsP^2}{k^2},  	 
   \frac{\QsA^2}{k^2}  	 
   \right)\, ,
     \label{Eq:SIPf}
 \end{align}
 where $\QsP$ and $\QsA$ are the saturation momenta for the projectile and target and $\alpha_s$ is the strong coupling constant. In the pioneering works of ~\cite{Krasnitz:1999wc,Krasnitz:2003jw,Lappi:2003bi,Blaizot:2010kh}, the function $f$ was studied numerically by solving Classical Yang-Mills (CYM) equations
 and projecting the classical field onto transversely polarized gluon states~(see e.g. Ref.~\cite{Schenke:2015aqa}); we will perform similar calculations in this work. So far the only known situation which is analytically tractable, is an expansion of $f$ in either one of its  arguments $\frac{\QsP^2}{k^2}$ and/or $\frac{\QsA^2}{k^2}$.
 In the {\it dilute-dense} approximation designed for asymmetric collision systems (e.g. p-A), one assumes 
 that for a given transverse momentum $k$ of interest the projectile is a dilute object, $  {\QsP^2}/{k^2} \lsim 1$. This allows for a systematic expansion of the production cross section in this parameter
\begin{align}
  \frac{dN}{d^2 k dy} = \frac{1}{\as} \, \left[
     \frac{\QsP^2}{k^2}
	  \ f_1 \! \left(  \frac{\QsA^2}{k^2} \right) +
    \left(  \frac{\QsP^2}{k^2} \right)^2 \ f_2 \! \left(  \frac{\QsA^2}{k^2} \right) + \ldots \right].
  \label{Eq:SIP}
\end{align}
Specifically, the function $f_1$ is known analytically for about two decades (see Refs.~\cite{Kovchegov:1998bi,Dumitru:2001ux,Blaizot:2004wu}); at this (leading) order the number of produced gluons for given projectile and target configurations is given by 
\begin{equation}
\left.\frac{dN}{d^{2}kdy}\right|_{\rho_{\rm p},\rho_{\rm t}}=\frac{2g^{2}}{(2\pi)^{3}}\int\frac{d^{2}q}{(2\pi)^{2}}\frac{d^{2}q'}{(2\pi)^{2}}\Gamma(\v{k},\v{q},\v{q'})\rho_{\rm p}^{a}(-\v{q'})\left[U^{\dagger}(\v{k}-\v{q'})U(\v{k}-\v{q})\right]_{ab}\rho_{\rm p}^{b}(\v{q}),
\label{Eq:SIPc}
\end{equation}
for a fixed configuration of color charges $\rho_{\rm p}$, $\rho_{\rm t}$ in the dilute projectile (p) and dense target (t). Here $\Gamma(\v{k},\v{q},\v{q' })$ is the square of Lipatov vertex, 
see \myref~\cite{Kovner:2018azs} or the main body of the paper for details. Although, equation~(\ref{Eq:SIPc}) is only quadratic in $\rho_{\rm p}$, it contains all orders of $\rho_{\rm t}$, which are re-summed in the adjoint Wilson line $U$, representing the eikonal scattering matrix for scattering of a single gluon on the target.

While Eq.~(\ref{Eq:SIPc}) provides the leading order in the dilute-dense expansion of Eq.~\eqref{Eq:SIP},
the function $f_2$, is also termed as the {\it first saturation correction} in the
projectile, since it comes along with two powers of $\QsP^2/k_\perp^2$,
corresponding to interactions with two valence sources in the projectile. 
Efforts to calculate $f_2$ analytically are detailed in~\myref\cite{Balitsky:2004rr} and more recently in~\myref\cite{Chirilli:2015tea}. However, at present, $f_2$ is known only partially.    

In summary, higher order corrections, functions $f_i$ for $i\ge 2$,
to the strict dilute-dense approximation, $f_1$, are presently
not known analytically.
Even if analytical forms of $f_i$ were known, they may still involve rather 
complicated momentum integrals 
of  Wilson lines of the target field, see e.g. Eq.~\eqref{Eq:SIPc}. 
Hence, it is practically inevitable to use numerical methods (typically involving lattice 
discretization) and we will therefore refer to this as a semi-analytical approach. 

Nevertheless, in contrast to CYM simulations, semi-analytical calculations based on $f_i$ neither require a numerical solution of the gauge field evolution in the forward light-cone nor a numerical implementation of 
LSZ reduction.  Besides, one additional advantage of this semi-analytic dilute-dense
approach is that it facilitates the inclusion of small-$x$ evolution, 
running coupling corrections, as well as higher order $\alpha_s$ corrections, in contrast to fully numerical CYM simulations. It is also superior in terms of simulation time and thus allows for an easier access to the continuum limit. However, the obvious drawback of this approach is that it may miss  a potentially large contribution from the  higher order expansion coefficients. Hence, the goal of this paper is to perform a systematic numerical study of the saturation corrections and to compare them with leading order dilute-dense approximation.

So far we have focused on single inclusive particle production; however an analogous discussion also applies to the multi-particle production. Specifically, the double inclusive two-gluon spectrum can be written in the following form, 
\begin{align}
	\frac{dN}{d^2 k_1 \, d y_1  \, d^2 k_2 dy_2} 
  = \frac{1}{\as^2} \, h
  \left( 
  \frac{\QsP^2}{k^2},  	 
   \frac{\QsA^2}{k^2}  	 
  \right)
    \label{Eq:DIPh}
 \end{align}
 with a new unknown function $h$. Here $k_1$ and $k_2$ are the gluons' transverse momenta and we assumed that $|k_1|=|k_2|=k$ to simplify notation. By considering a dilute projectile, we can again expand in ${\QsP^2}/{k_\perp^2}$, which results in 
\begin{align}
  \frac{dN}{d^2 k_1 \, d y_1 \, d^2 k_2 \, dy_2} =
  \frac{1}{\as^2} \, \left[ \left( \frac{\QsP^2}{k^2}  \right)^2 \ 
	  h_1
    \!\left(  \frac{\QsA^2}{k^2} \right) + \left( \frac{\QsP^2}{k^2}  \right)^3 \ h_2 \! \left( \frac{\QsA^2}{k^2}  \right) + \ldots
  \right] ,
  \label{Eq:DIP}
\end{align}
where the function $h_1$ can be found from the results of
\myrefs\cite{Kovner:2012jm,Kovchegov:2012nd,Altinoluk:2018ogz} and is also written in convenient form for numerical simulations  in \myrefs\cite{Kovner:2018fxj}. 
Compared to $f_1$ in Eq.~\eqref{Eq:SIPc}, which features two target Wilson lines (dipole), 
the function $h_1$ involves four Wilson lines (quadrupole). 
It is well known~\cite{Kovner:2012jm,Kovchegov:2012nd}, 
that this
part of the two-gluon production cross section is invariant  
under the reflection of either momenta $k_1$ or $k_2$ and thus 
generates only even harmonics of azimuthal anisotropy. 
More recently, in \myrefs~\cite{McLerran:2016snu,Kovchegov:2018jun},
it was shown that this accidental symmetry with respect to the 
reflection of one of the momenta is lifted by the 
first saturation contribution, $h_2$, to double inclusive production. 
In particular, the part of $h_2$ responsible for the
odd harmonics was derived analytically~\cite{McLerran:2016snu,Kovchegov:2018jun}. 
Nevertheless, the full result for $h_2$ (including also the first saturation corrections to the even part) is currently unknown and would require determination of $f_2$. 

In the current study, we will use semi-analytical results for $f_1$, $h_1$ and the odd part of $h_2$ in order to compute the observables $dN/dy$, $v_2$, $v_4$ and $v_{3}$ which are of particular relevance to phenomenological CGC studies~\cite{Schenke:2015aqa,Schenke:2016lrs,Kovner:2018azs,Kovner:2018fxj, Mace:2018vwq,Mace:2018yvl,Mace:2019rtt}. By explicitly comparing the results for these observables with the corresponding ones obtained in full dense-dense calculations of Eqs.~\eqref{Eq:SIPf} and~\eqref{Eq:DIPh} based on CYM simulations, we will assess the quality of the dilute-dense approximation and the impact of higher order saturation corrections.

\section{Dilute-dense vs. Dense-Dense -- Explicit results and comparison} 
\subsection{General Setup}
We consider the color charge distribution in the dilute projectile as
\begin{eqnarray}
\label{eq:RhoPModel}
\langle \rho_{a}^{(p)}(\v{x}) \rho_{b}^{(p)}(\v{x'}) \rangle = \left(\frac{g^2\mu}{Q_s}\right)^2~\QsP^2\left(\frac{\v{x}+\v{x'}}{2}\right)~\delta_{ab}~\delta^{(2)}(\v{x}-\v{x'})\;,
\end{eqnarray}
where the local saturation scale $\QsP^{2}\left(\v{x}\right)$ is determined by
\begin{eqnarray}
\QsP^2\left(\v{x}\right)= 2\pi R_p^2~T(\v{x})~(Q_{s,0}^{(p)})^2= (Q_{s,0}^{(p)})^2~\exp\left( -\frac{\v{x}^{\,2}}{2R_{p}^2} \right)
\end{eqnarray}
such that the dilute projectile can be thought of as a minimal saturation model for the proton. We note that in order to scrutinize the particle production mechanism we restrict ourselves to such a minimal model, and have not included additional ingredients, such as e.g. sub-nucleonic constituents \cite{Schlichting:2014ipa,Mantysaari:2017cni} or saturation scale ($Q_{s,0}^{(p)}$--) fluctuations \cite{McLerran:2015qxa} commonly invoked in phenomenological CGC calculations. Similarly, we consider a spatially homogeneous color charge distribution of the dense target, i.e.
\begin{eqnarray}
\label{eq:RhoTModel}
\langle \rho_{a}^{(t)}(\v{x}) \rho_{b}^{(t)}(\v{x}') \rangle = \left(\frac{g^2\mu}{Q_s}\right)^2~(Q_{s,0}^{(A)})^2~\delta_{ab}~\delta^{(2)}(\v{x}-\v{x'})
\end{eqnarray}
which likewise, can be thought of as a simplistic saturation model of a very large nucleus. We note that the parameter $(Q_{s,0}^{(A)})^2$ characterizes the saturation scale everywhere in the large nucleus, whereas $(Q_{s,0}^{2 (p)})^2$ gives the saturation scale of the proton in the center, such that on average the saturation momentum in the proton is somewhat smaller, e.g. $\langle Q_{s}^{2}(\v{x})\rangle|_{|\v{x}|<R_p} \approx 0.79~(Q_{s,0}^{(p)})^2$.  

Since the common prefactor $\left(\frac{g^2\mu}{Q_s}\right)$ can always be absorbed into a redefinition of the saturation momenta $Q_{s,0}^{(p/A)}$, we will always fix its value to $\left(\frac{g^2\mu}{Q_s}\right)=1.42857$ -- a typical value employed in phenomenological studies in IP-Glasma~\cite{Schenke:2012wb,Schenke:2012fw}. Besides the projectile size $R_p$ and the saturation momenta $Q_{s,0}^{(p/A)}$ of the projectile and target, our model then only has one additional parameter $m$ which regulates the infrared behavior of the color charge distributions. We again follow previous works~\cite{Schenke:2012wb,Schenke:2012fw}, and adopt the common procedure to replace 
\begin{equation}
\rho_{a}(\v{p}) \to \frac{\v{p}^{\,2}}{\v{p}^{\,2}+m^2}  \rho_{a}(\v{p}) 
\end{equation}
in our numerical calculations, where evidently $Q_{s,0}^{2(p/A)} \gg m^2$ should hold to minimize the sensitivity to the infrared regulator. If not stated otherwise, we will fix $R_p=2$~${\rm GeV}^{-1}$ and $m=0.5$~GeV  in the following and compare results for particle production in dilute-dense and dense-dense calculations as a function of the saturation scales $Q_{s,0}^{(p)}$ and $Q_{s,0}^{(A)}$ of the projectile and target.\footnote{Stated differently we set the dimensionless parameter $m R_p=1$ and use $R_p=2\ {\rm GeV}^{-1}$ to set the scale in our calculation.} Details of the individual calculations proceed as described as outlined below, and are described in detail in \myref\cite{Mace:2019rtt} for dilute-dense and \myref\cite{Schenke:2015aqa} for dense-dense calculations. 

\subsection{Dilute-dense approximation}
We briefly review the {\bf dilute-dense} expressions derived previously in \myrefs\cite{Kovchegov:1998bi,Dumitru:2001ux,Blaizot:2004wu,Kovner:2012jm,Kovchegov:2012nd,McLerran:2016snu,Kovchegov:2018jun}). Starting from a statistical sampling of the color charge distribution $\rho_{p}$ and $\rho_{t}$ of the projectile and target according to Eqns.~(\ref{eq:RhoPModel},\ref{eq:RhoTModel}), the leading contribution to the configuration-by-configuration spectrum is then given by 
Eq.~\eqref{Eq:SIPc} with 
\begin{equation}
    \Gamma(\v{k},\v{q},\v{q'}) = 
    \left(\frac{\v{k}}{k^2} -  \frac{\v{q}}{q^2} \right) \cdot 
     \left( \frac{\v{k}}{k^2} -  \frac{\v{q'}}{{q'}^2}  \right). 
\end{equation}
However, as alluded to in the Sec.~\ref{sec:Intro}, the form  \eqref{Eq:SIPc}  is, not particularly useful for numerical calculations, as it involves two two-dimensional, not obviously factorizable integrals.   Instead one can recast  \eqref{Eq:SIPc}  in fully equivalent form 
\begin{align}
	\label{Eq:even}
	\frac{d N^{\rm even} (\v{k})}{d^2k dy} \Big[\rho_p, \rho_t\Big] &=  \frac{2}{(2\pi)^3}
		\frac{ \delta_{ij} \delta_{lm}  +  \epsilon_{ij} \epsilon_{lm} }{k^2}
		\,\Omega^a_{ij} (\v {k})
		\left[ \Omega^a_{lm} (\v {k}) \right]^\star
\end{align}
with $\Omega(\v{k})$ defined as the Fourier transform of 
\begin{equation}
	\Omega_{ij}^a(\v{x}) =
	g 
	\left[\frac{\partial_i }{\partial^2}
	\rho^b_p(\v{x})
	\right]
	\partial_j U_{ab} (\v{x})\,,
	\label{Eq:Omega}
\end{equation}
and $\epsilon_{ij} (\delta_{ij})$ denotes the Levi-Civita symbol (Kronecker delta). The adjoint Wilson line $U_{ab}$ is a functional of the target charge density: 
\begin{equation}
U(\v{x}) = {\cal P} \exp \left( i g^2 \int dx^+ \frac{1}{\partial^2} \rho_t^a(x^+, \v{x}) T_a \right). 
	\label{Eq:U}
\end{equation}
The expression in Eq.~(\ref{Eq:even}) explicitly demonstrates that numerical evaluation of the even component boils down to a straightforward computation of two combination $\epsilon_{ij} \Omega^a_ij(\v{x})$ and $\delta_{ij} \Omega^a_{ij}(\v{x})$ complemented  by  the fast Fourier transformation.  
As discussed in Introduction,  the 
odd component of the particle production cross section is given by higher order corrections to the strict dilute-dense limit. The leading contribution is 
\begin{align}
	\label{Eq:odd}
	  &\frac{d N^{\rm odd} (\v{k})}{d^2 k dy} \Big[\rho_p, \rho_t  \Big]
    =
	{ \frac{2}{(2\pi)^3} }
	{\rm Im}
	\Bigg\{
		\frac{g}{{\v{ k}}^2}
		\int \frac{d^2 l}{(2\pi)^2}
				\frac{  {\rm Sign}({\v{k}\times \v{l}}) }{l^2 |\v{k}-\v{l}|^2 }
		f^{abc}
			\Omega^a_{ij} (\v{l})
			\Omega^b_{mn} (\v{k}-\v{l})
			\left[\Omega^{c}_{rp} (\v{k})\right]^\star \notag
		 \\ &  \times \quad
		\left[
			\left(
			{\v{ k}}^2 \epsilon^{ij} \epsilon^{mn}
		-\v{l} \cdot (\v{k} - \v{l} )
		(\epsilon^{ij} \epsilon^{mn}+\delta^{ij} \delta^{mn})
		\right) \epsilon^{rp}+
		2 \v{k} \cdot (\v{k}-\v{l}) { \epsilon^{ij} \delta^{mn}} \delta^{rp}
		\right]
	\Bigg\} \, .
\end{align}
We emphasize that in a Gaussian model for the projectile the above expression does {\it not} contribute to single inclusive cross section due to $\left \langle \frac{d N^{\rm odd} (\v{k})}{d^2 k dy} [\rho_p, \rho_t  ]  \right \rangle  = 0 $, but does contribute to the double inclusive production.

Based on the configuration-by-configuration spectrum, the single and double inclusive gluon production cross section are then given in terms of the statistical averages of the respective operators
\begin{align}
\label{eq:SingleInclusiveDD}
    \frac{d N (\v{k})}{d^2 k dy} &=
    \left \langle \frac{d N (\v{k})}{d^2 k dy} \Big[\rho_p, \rho_t  \Big] \right \rangle \, , \\ 
\label{eq:DoubleInclusiveDD}
    \frac{d^2 N (\v{k_1}, \v{k_2})}{d^2 k_1 dy_1 d^2 k_2 dy_2  } &=
    \left \langle 
    \frac{d N (\v{k_1})}{d^2 k_1 dy_1} \Big[\rho_p, \rho_t  \Big] 
    \frac{d N (\v{k_2})}{d^2 k_2 dy_2} \Big[\rho_p, \rho_t  \Big] \right \rangle
\end{align}
where brackets $\langle . \rangle$ denote the statistical average over different realizations of the color charge configurations of the projectile and target, and
\begin{equation}
    \frac{d N (\v{k})}{d^2 k dy} \Big[\rho_p, \rho_t  \Big]  = 
    \frac{d N^{\rm even} (\v{k})}{d^2 k dy} \Big[\rho_p, \rho_t  \Big]
    +\frac{d N^{\rm odd} (\v{k})}{d^2 k dy} \Big[\rho_p, \rho_t  \Big]\,. 
\end{equation}
We note that Eq.~(\ref{eq:DoubleInclusiveDD}) relies on the fact that to leading order in $\alpha_s$, the two-particle correlation function $\left.\frac{dN}{dy_1 d^2p_1 dy_2 d^2p_2}\right|_{\rho_{p},\rho_{t}}=\left.\frac{dN}{dy_1 d^2p_1} \frac{dN}{dy_2 d^2p_2}\right|_{\rho_{p},\rho_{t}}$ factorizes into a product of single particle distribution when evaluated for a fixed configuration of color charges of the projectile and target~\cite{Gelis:2008ad,Gelis:2008sz,Lappi:2009xa,Schenke:2015aqa}. Genuinely non-factorizable (``non-flow'') two-particle correlations e.g. due to di-jet production only appear at next-to-leading order in $\alpha_s$; however they have not been calculated in the dense-dense limit so far.

\subsection{Dense-dense CYM}
Notably the \textbf{dense-dense calculation} proceeds along very similar lines. Starting from the statistical sampling color charges in the projectile and target, one computes the light like Wilson lines of both projectile and target in fundamental representation. Subsequently, the initial gauge fields in the forward light-cone (at $\tau=0^{+})$ are determined according to the solution of the classical Yang-Mills equations, but now including the full non-linearity of projectile and target fields. Starting from these initial conditions, one then solves the classical Yang-Mills equations of motion in the forward light cone up to the time where observables are measured. Ultimately, one exploits the residual gauge freedom to fix Coulomb gauge at the time of the measurement and determines the spectrum of produced gluons $\frac{dN_{g}}{dy d\v{p}}$  by projecting the gauge fields onto transversely polarized modes (c.f \cite{Berges:2013fga}). Details of the numerical procedure can be found in \myref\cite{Schenke:2015aqa}.

\subsection{Numerical results}

\begin{figure}
\begin{minipage}{0.475\textwidth}
\includegraphics[width=\textwidth]{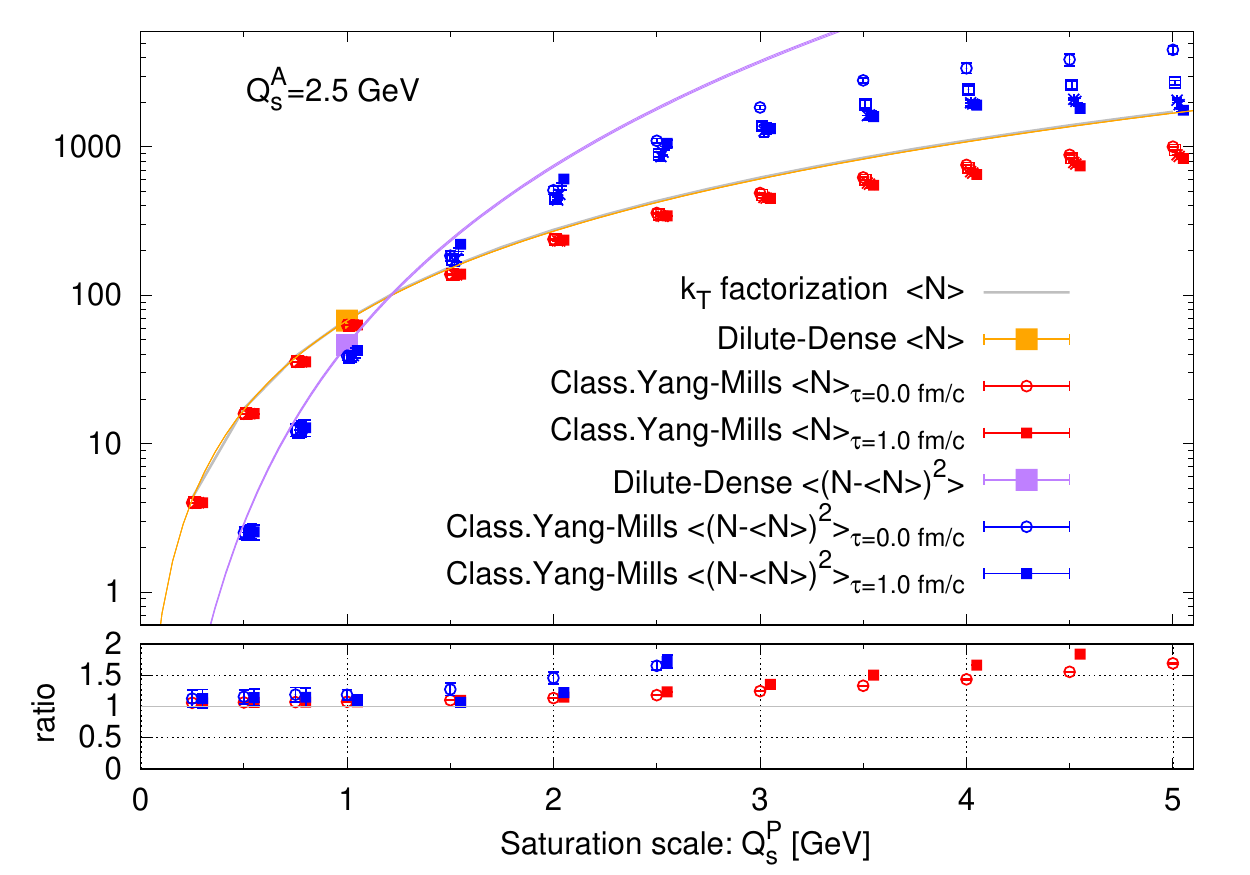}
\end{minipage}
\begin{minipage}{0.475\textwidth}
\includegraphics[width=\textwidth]{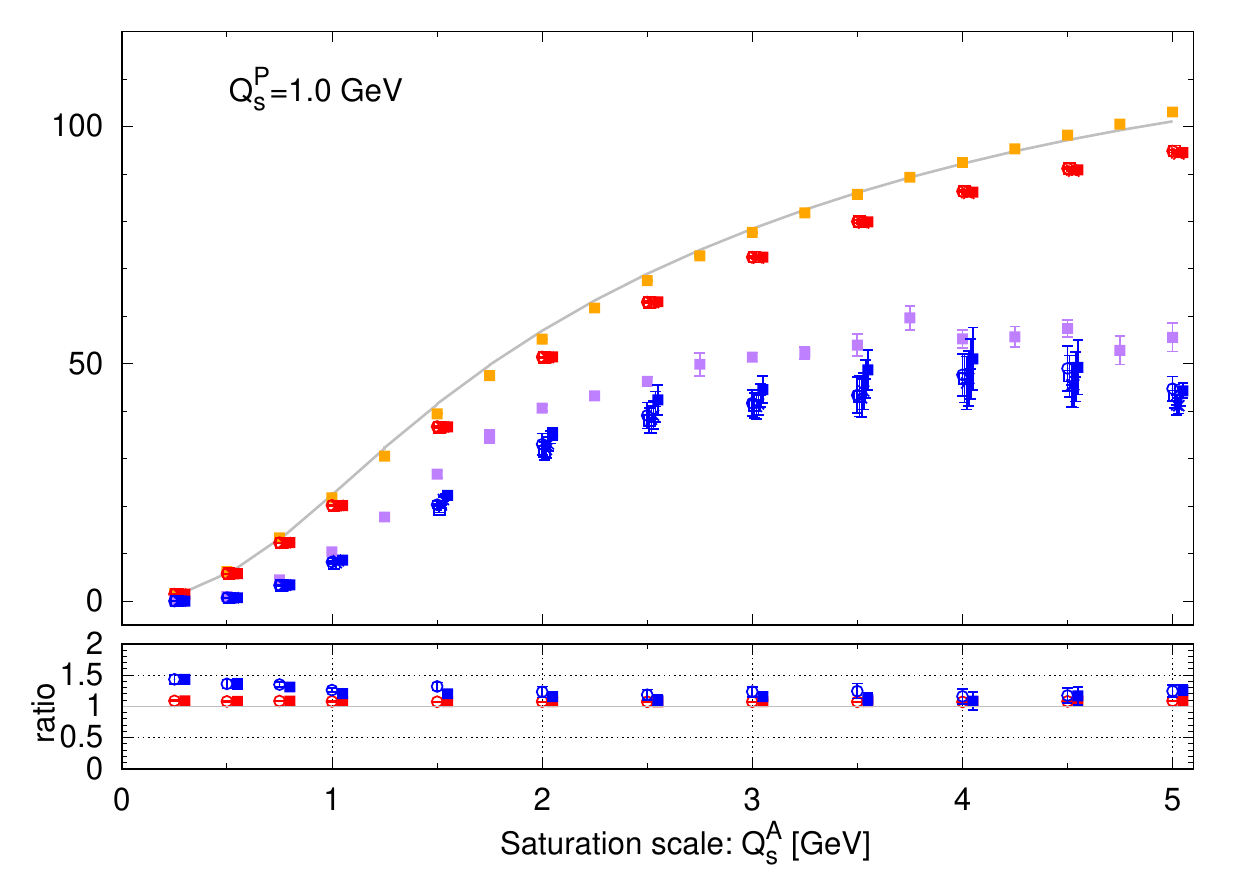}
\end{minipage}
\caption{\label{fig:N} Single inclusive multiplicity $\langle dN/dy\rangle$ and its variance (left) as a function of $Q_{s,0}^{(p)}$ and (right) as a function of $Q_{s,0}^{(A)}$. Different curves show results obtained from dilute-dense and dense-dense (Class.~Yang-Mills) calculations. Gray line show an additional semi-analytic calculation based on $k_t$-factorization~\cite{Kovchegov:1998bi,Dumitru:2001ux,Blaizot:2004wu}, which is equivalent to the leading order dilute-dense approximation for single inclusive production. Classical Yang-Mills results obtained for different times $\tau=0.0,0.2,0.4,0.6,0.8,1.0\, {\rm fm}/c$ are shown as different symbols and have been offset horizontally for better visibility. Bottom panels show the ratio of dilute-dense to dense-dense at $\tau=0$ (open circles) and $\tau=1\ {\rm fm}/c$ (full squares).}
\end{figure}

We now turn to the discussion of our numerical results and first study the overall multiplicity per unity rapidit $\left\langle \frac{dN_{g}}{dy} \right\rangle \equiv\int_{m/2}d^2\v{p} \frac{dN_{g}}{dy d^2\v{p}}$
and its variance $\left\langle \frac{dN_{g}}{dy} \frac{dN_{g}}{dy} \right\rangle - \left\langle \frac{dN_{g}}{dy} \right\rangle^2$ which are presented in Fig.~\ref{fig:N}. We first focus on the left panel, where we fix the saturation scale of the target $\QsA=2.5\  {\rm GeV}$ and investigate the dependence on the saturation scale $\QsP$ of the projectile, which as discussed in Sec.~\ref{sec:Intro}, corresponds to the expansion parameter for the dilute-dense approximation. Since the multiplicity in dilute-dense calculations is directly proportional to $\left(\QsP\right)^2$ it is in fact sufficient to perform a single set of calculations shown as the single data point for $\QsP=1\ {\rm GeV}$. Based on the analytical scaling $\langle dN/dy\rangle \propto \left(\QsP\right)^2$ of the dilute-dense formulae for the single inclusive production and $\left(\QsP\right)^4$ for the variance, one then obtains the result for all other values of $\QsP$ shown in terms of the solid lines. By comparing to the data-points of the full dense-dense (CYM) calculation, we find that such scaling is indeed well reproduced for $\QsP \lesssim 1\ {\rm GeV}$, where there is a good quantitative agreement between dilute-dense and dense-dense calculations. 

However, for $\QsP \gtrsim 1 \ {\rm GeV}$ higher order saturation corrections become increasingly important and the dilute-dense approximation tends to over-predict particle production. While for single inclusive particle production higher order corrections remain on the order of $\sim 50\%$ even up to $\QsP\sim 4\ {\rm GeV}$, we find that double inclusive observables, such as the variance of the multiplicity, appear to be significantly more sensitive to higher-order saturation corrections. When considering $\left\langle \frac{dN_{g}}{dy} \frac{dN_{g}}{dy} \right\rangle - \left\langle \frac{dN_{g}}{dy} \right\rangle^2$, sizeable discrepancies on the order of $\sim 50\%$ between dilute-dense and dense-dense calculations already emerge for $\QsP \sim 2 ~{\rm GeV}$ and steadily increase for larger values of $\QsP$ where the dilute-dense approximation breaks down. 

By fixing the saturation scale of the projectile $\QsP$ to a (small) value of $1\ {\rm GeV}$  we can further assess the $\QsA$ dependence of the dilute-dense approximation. We find that for such relatively small values of $\QsP$, the dilute-dense approximation provides a more or less uniform approximation of the dense-dense result, as is shown in the right panel of Fig.~\ref{fig:N}. Higher order saturation corrections to $\left\langle \frac{dN_{g}}{dy} \right\rangle$ and it's variance are typically $\lesssim 20\%$ except perhaps for very small values of $\QsA \lesssim 1\ {\rm GeV}$, where the projectile effectively becomes more dense than the target. We also note that for the dense-dense calculation changes in the overall multiplicity as well as its fluctuations are relatively small over the course of the classical Yang-Mills evolution in the forward light-cone, indicating that the values inferred at $\tau=0^{+}$ already provide a good estimate of the event-by-event gluon multiplicity.

Next we investigate the effects on azimuthal correlations between the produced gluons, which we will quantify in terms of Fourier coefficients $v_{n}\{2\}$ of the $\v{p}$ integrated two-particle correlation function
\begin{eqnarray}
v_{n}\{2\}=\sqrt{\frac{\langle b_{n}b_{n}^{*} \rangle}{\langle b_{0}b_{0}^{*}\rangle}}\;, \qquad b_{n}\equiv\int_{m/2}d^2\v{p} \frac{dN_{g}}{dy d^2\v{p} }~e^{in \phi_{\v{p}}}\;.
\end{eqnarray}
Such initial state azimuthal correlations reflect intrinsic correlations of gluons in the projectile and target \cite{Lappi:2015vta,Altinoluk:2015uaa}; they are currently of particular phenomenological interest, as various studies have argued for their importance in understanding collective phenomena in small collision systems \cite{Schenke:2016lrs,Greif:2017bnr,Mace:2018vwq,Mace:2018yvl,Mace:2019rtt}, including p/d/He3+A collisions at RHIC as well as p+p and p+A collisions at the LHC.

\begin{figure}
\begin{minipage}{0.475\textwidth}
\includegraphics[width=\textwidth]{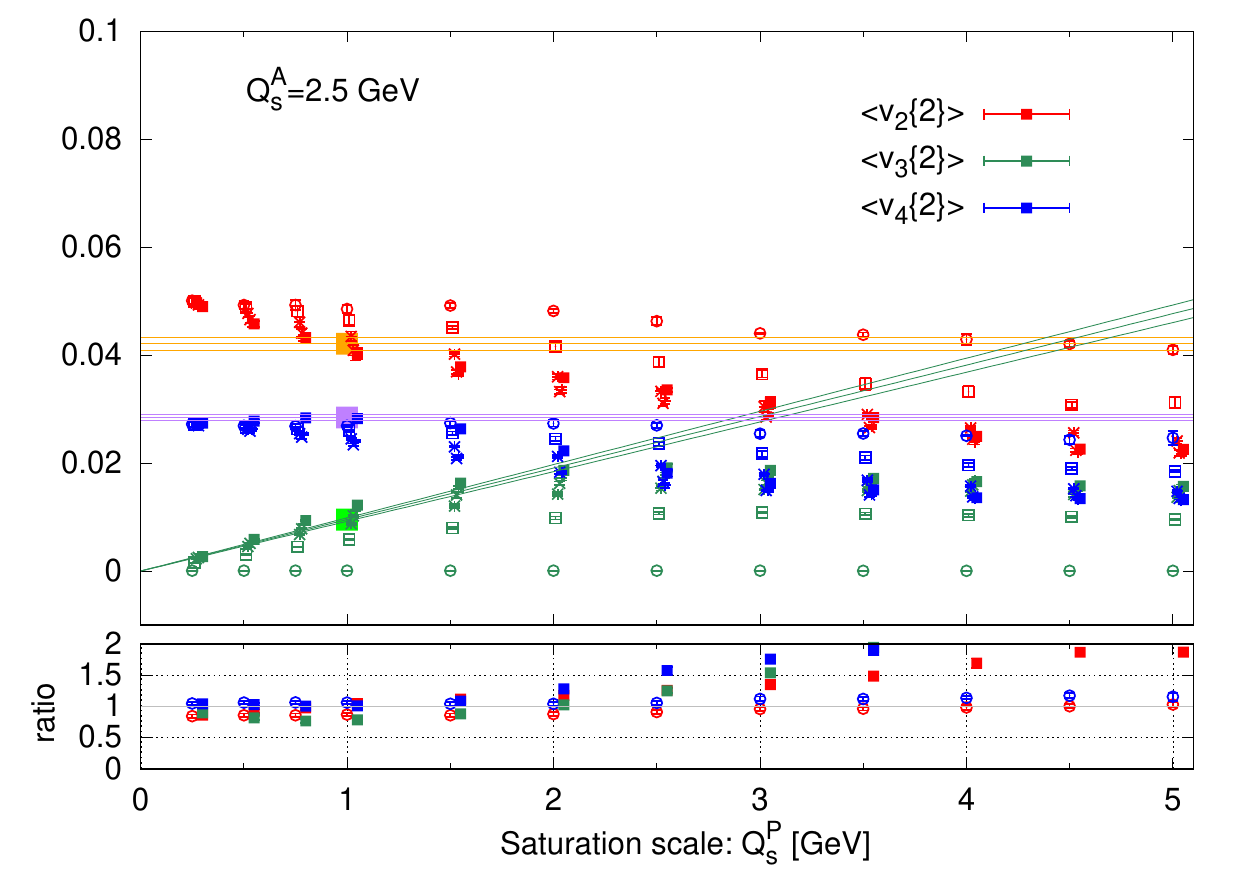}
\end{minipage}
\begin{minipage}{0.475\textwidth}
\includegraphics[width=\textwidth]{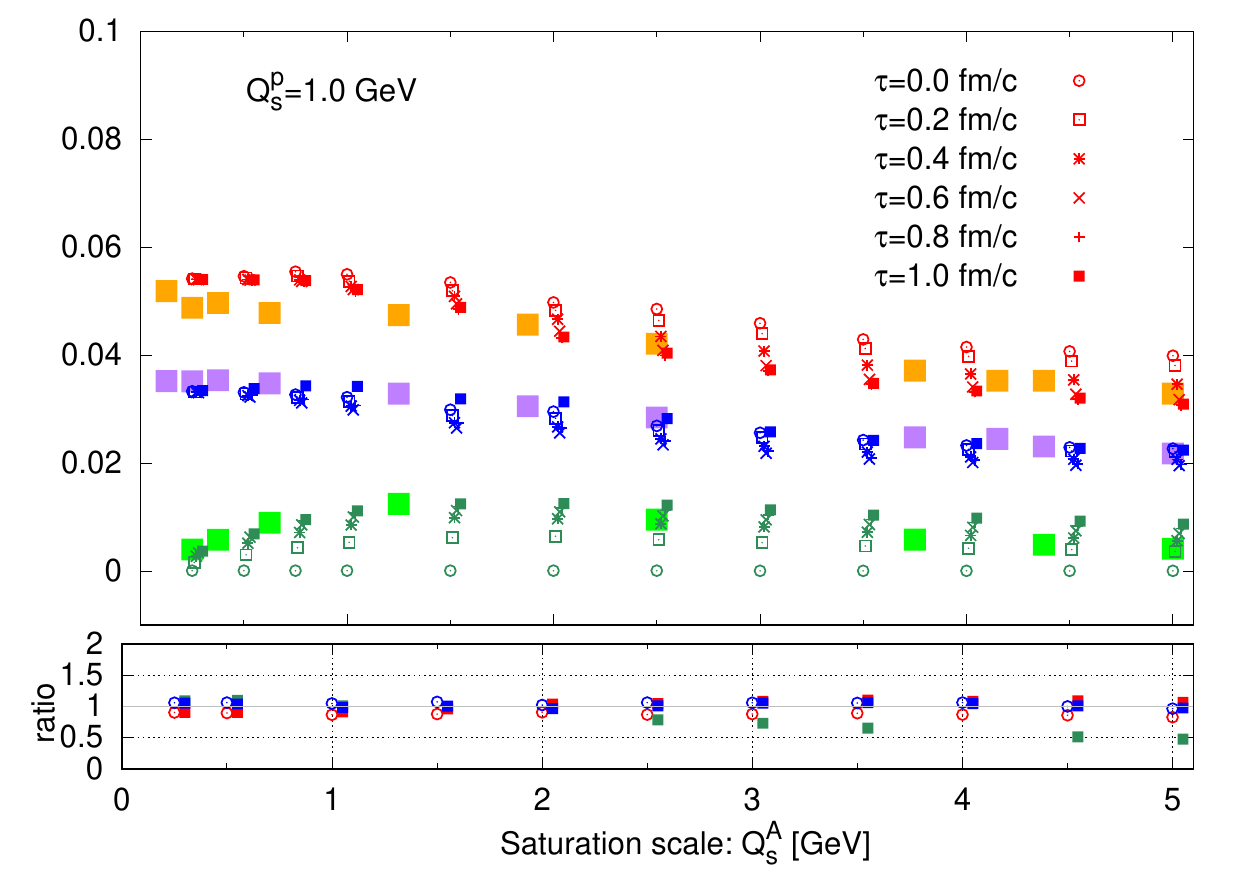}
\end{minipage}
\caption{\label{fig:vN} Azimuthal anisotropy $v_{n}\{2\}$ (left) a function of $Q_{s,0}^{(p)}$ and (right) as a function of $Q_{s,0}^{(A)}$. Different curves show results obtained from dilute-dense and dense-dense (Class. Yang-Mills) calculations. Classical Yang-Mills results obtained for different times $\tau=0.0,0.2,0.4,0.6,0.8,1.0 \ {\rm fm}/c$ are shown as different symbols and have been offset horizontally for better visibility. Bottom panels show the ratio of dilute-dense to dense-dense at $\tau=0$ (open circles) and $\tau=1 \ {\rm fm}/c$ (full squares). Solid lines in the left panel are obtained using the scaling argument of Ref.~\cite{Mace:2018yvl} and represent $v_{2},v_{4} \propto (\QsP)^0$ and $v_3 \propto  \QsP $. }
\end{figure}

We present a compact summary of our results for azimuthal correlations in Fig.~\ref{fig:vN}, where we compare the results for $v_{2}$,$v_{3}$ and $v_{4}$ in dilute-dense and dense-dense calculations as a function of $\QsP$ and $\QsA$ in the left and right panels. While for small values of $\QsP$, the 
dense-dense calculation appears to well approximated by the semi-analytic dilute-dense calculation, and shows the expected scaling of the different harmonics $v_{2},v_{4} \propto (\QsP)^0$ and $v_3 \propto  \QsP $~\cite{Mace:2018yvl}, the dilute-dense approximation starts to overpredict the azimuthal correlation strength for $\QsP \gtrsim 1\  {\rm GeV}$. We find that in this regime, the $v_{n}$'s obtained in the dense-dense calculation start to show a significant time dependence, clearly indicating the importance of re-scattering in the forward light cone (in the ``final state''), which are always associated with saturation corrections to the leading order dilute dense result. Specifically, the odd-harmonic $v_{3}$ increases from $v_{3}=0$ at $\tau=0^{+}$ up to $v_{3}\approx 2 \%$ over the course of the classical Yang-Mills evolution, while the even harmonics $v_{2}$ and $v_{4}$ decrease by a comparable amount (see also \cite{Schenke:2015aqa}). Since for $\QsP \lesssim 2$ GeV the increase of $v_{3}$ is rather well described by the first saturation correction to dilute-dense limit (which gives zero $v_3$), it is conceivable that the observed decrease of the even harmonics could also be captured (at least partially) by the first saturation correction. Indeed, naive power counting argument in $\QsP$
predicts linear deviation of $v_{\rm even}$ from the dilute-dense regime. However, it is impossible to make this argument stronger, because, as discussed in Sec.~\ref{sec:Intro}, the associated corrections to $d^2N_{\rm even}/d^2k_1 d^2k_2$ have not been calculated to date.

We also observe from the left panel of Fig.~~\ref{fig:vN} that in order to obtain the phenomenologically relevant ordering of the different harmonics $v_{2}>v_{3}>v_{4}$, one needs to access relatively large values of $\QsP$ which appear to be outside the range of validity of the dilute-dense approximation (as demonstrated e.g. by $v_3$ deviating from its linear growth at lower $\QsP$). Vice versa for the relatively small values of $\QsP=1\ {\rm GeV}$ shown in the right panel, the even harmonics $v_{2}$ and $v_{4}$ exhibit a significantly smaller time dependence, resulting in $v_{4}>v_{3}$ irrespective of the value of $\QsA$. We find that in this regime, where the dilute-dense approximation is justified, the $\QsA$ dependence of the dense-dense calculation is indeed well reproduced by the semi-analytic dilute-dense calculation, with typical errors on the $10\%$ level for. This is, however, not the case for $v_3$ which is underpredicted by the dilute-dense approximation at large values of $\QsA$.

\section{Conclusions} 
In this article, we presented for the first time a numerical study of  higher order saturation corrections to the leading order dilute-dense approximation for different phenomenologically relevant observables.  We explicitly demonstrated the expected deviations between the dilute-dense approximation and full dense-dense CYM simulations for single and double inclusive observables and showed that these deviation increase with the saturation momentum of the dilute projectile, as we anticipated based on the expansion~\eqref{Eq:SIPf}. While for single inclusive observables, such as the average multiplicity $\langle dN_{g}/dy \rangle$, we find that the deviations remain on the order of $50\%$ even when the saturation scale of the projectile becomes on the same order as the saturation scale of the target ($\QsP \sim \QsA$), we find that double inclusive observables such as the azimuthal correlations $v_{n}$ are significantly more sensitive to higher order saturation corrections.

When restricted to the range of validity, i.e. for $\QsP \ll \QsA$, we find that the dilute-dense approximation faithfully reproduces the dense-dense results, with almost uniform accuracy as a function of the saturation scale of the target $\QsA$. We find that in this regime, the dilute-dense approximation tends to over-predict particle production only by about $10 (20)\%$ compared to the corresponding dense-dense result for the single (double) inclusive gluon production.

\textit{Acknowledgements:} We thank A.~Dumitru, A.~Kovner, Yu.~Kovchegov,  T.~Lappi, M.~Mace, L.~McLerran, B.~Schenke,  P.~Tribedy and R.~Venugopalan for insightful discussions and collaboration on related projects. We gratefully acknowledge support by the Deutsche Forschungsgemeinschaft (DFG) through the grant CRC-TR 211 “Strong-interaction matter under extreme conditions” Project number 315477589 (S.S.) and by the US Department of Energy grant DE-SC0020081 (V.S.). V.S. thanks the ExtreMe Matter Institute EMMI (GSI Helmholtzzentrum f\"ur Schwerionenforschung, Darmstadt, Germany) for partial support and hospitality.

 \bibliography{bibl}
 
\end{document}